\documentclass[prl,amsmath,amssymb,dvips,twocolumn,showpacs]{revtex4}

\usepackage{graphics,epsfig,amssymb}

\def\beq{\begin{equation}} 
\def\eeq{\end{equation}} 
\begin{document}

\title{Four-body correlations  in  nuclei}

\author{M. Sambataro$^a$ and N. Sandulescu$^b$}
\affiliation{$^a$Istituto Nazionale di Fisica Nucleare - Sezione di Catania,
Via S. Sofia 64, I-95123 Catania, Italy \\
$^b$National Institute of Physics and Nuclear Engineering, P.O. Box MG-6, 
Magurele, Bucharest, Romania}

\begin{abstract}
Low-energy spectra of 4$n$ nuclei are described with high accuracy  in terms of  four-body correlated structures (``quartets"). The states of all $N\geq Z$ nuclei belonging to the $A=24$ isobaric chain are represented as a
superposition of two-quartet states, with quartets being characterized by isospin $T$ and angular momentum $J$. These quartets are assumed to be 
those describing the lowest states in $^{20}$Ne ($T_z$=0),  $^{20}$F ($T_z$=1) and $^{20}$O ($T_z$=2). We find that
 the spectrum of the self-conjugate nucleus $^{24}$Mg can be well reproduced in terms of $T$=0 quartets only
and that, among these, the $J$=0  quartet plays by far the leading role in the structure of the ground state. The same conclusion is drawn in the case of the three-quartet $N=Z$ nucleus $^{28}$Si. As an application of the quartet formalism to nuclei not confined to the $sd$ shell, we provide a description of the low-lying spectrum of the proton-rich
$^{92}$Pd. The results achieved indicate that, in 4$n$ nuclei, four-body degrees of freedom are more important and more general than  usually expected.  

\end{abstract}

\maketitle
In  nuclear physics, four-body correlated structures are usually associated with $\alpha$-clustering. $\alpha$-clustering is known to be a relevant phenomenon in light $N=Z$ nuclei
especially at excitation energies close to the $\alpha$ emission threshold. A common theoretical approach to $\alpha$-clustering is represented by the $\alpha$-cluster model \cite{ikeda}. According to this model, the nucleus consists of a $N=Z$ core to which some $\alpha$ clusters are appended. These clusters are
tightly bound and spatially localized structures of two neutrons
and two protons. The $\alpha$-cluster  model exhibits
 a striking contrast with the standard shell model picture in which protons and neutrons are described instead as weakly interacting quasiparticles in a mean-field. However,
these two pictures are expected to coexist at low excitation energies where,
due to the Pauli blocking which acts stronger than in the states close to
the $\alpha$ emission  threshold (e.g.,  see the case of the Hoyle state
\cite{hoyle}),
the $\alpha$-clustering is expected  to manifest itself mainly as  four-body correlations
 in the configuration space.
  It is thus commonly supposed that
  correlated  four-body structures (``quartets") play a major role in the ground and excited sates of $N=Z$ nuclei.
    However, to our knowledge, this supposition has never been supported by compelling calculations. 
    In  this letter, by using a simple microscopic quartet model, we will show how the low-energy 
    spectra of 4$n$ nuclei can be indeed described in terms of quartets  with an accuracy comparable with 
     state-of-the-art shell model calculations. 
     We shall  prove that this is the case not only for self-conjugate  nuclei but also for 4$n$ nuclei  with $N \neq Z$. 
      For the latter nuclei the low-lying states will be expressed  in terms  of quartets built not only by  two protons and two 
       neutrons but also by one proton and three neutrons  as well as by four neutrons. This fact indicates that  the four-body degrees
       of freedom are important also for configurations which are different from the $\alpha$-like ones.
       
   Microscopic quartet models have a long history in nuclear structure. They have been employed to treat 
   the proton-neutron ($pn$) interaction, in particular the $pn$ pairing, and to investigate the quartet condensation
   in $N=Z$ nuclei \cite{flowers,arima,yamamura,dobes,hasegawa,zelevinsky}. However, their complexity has always represented 
   a hindrance to their development.  Recently, a  simple approach has been proposed which is able to describe
   accurately the ground state of the  isovector pairing Hamiltonian as a condensate of quartets built by two neutrons and two protons
   coupled to total isospin $T=0$ and, for spherically symmetric mean fields, to total angular momentum $J$=0 \cite{qcm}. This quartet model conserves exactly the particle number, the isospin and the Pauli principle and 
   can be applied for any number of quartets.   Later on a more general description of the  ground state of $N=Z$ 
   nuclei as a product of distinct quartets has been proposed and applied to a description of both isovector ($T=1$, $J=0$) and isoscalar  ($T=0$, $J=1$) pairing correlations \cite{sasat1,sasaj0}.
   
   Of course, pairing is only a part (although a crucial one) of the nuclear interaction. 
It is reasonable to expect that a realistic
    description of self-conjugate nuclei should involve not only $T=0$, $J=0$ quartets.
In the following we will show how the approach of Refs. \cite{sasat1,sasaj0} has been  extended to include quartets 
with arbitrary values of isospin and angular momentum and to treat realistic interactions of shell model type.
 
One of the  motivations of the present work has been that of identifying the quartets which contribute most to the structure of the low-lying states in even-even $N=Z$ nuclei and, in particular, of investigating the role played by $T=0$, $J=0$ quartets in the ground state of these nuclei. Our analysis will be mainly confined to $sd$-shell nuclei. We will begin by fixing a set of $T=0, 1, 2$ quartets and then use these quartets for a description of the low-lying states of $^{24}$Mg. To such a purpose, we will carry out configuration interaction calculations in spaces built in terms of these quartets. By progressively restricting this set of quartets we will identify the most relevant ones for the low-lying states of this nucleus and, in particular, for the ground state. A confirmation of the conclusions reached for $^{24}$Mg will then be searched in the case of $^{28}$Si. 
 Having defined a set of $T=0$, 1, 2 quartets, a large variety of nuclear systems become, in principle, within reach in addition to self-conjugate nuclei.
 These include both even-even and odd-odd $4n$ nuclei.  In the second part of the paper, we will provide a description of the whole isobaric chain of $A=24$ nuclei with $N>Z$ in the formalism of these quartets. Finally, as an application of the formalism to nuclei not confined to the $sd$ shell, we will discuss the case of the proton-rich $^{92}$Pd.

We start by briefly illustrating our approach. We work in a spherically symmetric mean 
field and label the single-particle states by
$i\equiv \{n_i,l_i,j_i\}$, where the standard notation for the orbital quantum numbers is used. The  quartet creation operator is defined as
\begin{eqnarray}
Q^+_{\alpha ,JM,TT_z}&=&\sum_{i_1j_1J_1T_1}\sum_{i_2j_2J_2T_2}
C^{(\alpha )}_{i_1j_1J_1T_1,i_2j_2J_2T_2}\nonumber\\
&&\times\Bigl[ [a^+_{i_1}a^+_{j_1}]^{J_1T_1}[a^+_{i_2}a^+_{j_2}]^{J_2T_2}\Bigr]^{JT}_{MT_z}\nonumber,
\end{eqnarray}
where $a^+_i$ creates a fermion in the single particle state $i$ and
$J(T)$ and $M(T_z)$ denote, respectively, the total angular momentum(isospin) and the relative projections. No restrictions on the intermediate couplings $J_1T_1$ and $J_2T_2$ are introduced in the calculations. 
In order to generate the spectra of $4n$ nuclei we perform configuration interaction calculations in spaces built in terms of selected sets of the above quartets. For these calculations
we adopt a $static$ definition of the quartets, i.e. after selecting the quartets (according to a criterion specified below) we keep them ``frozen" in all calculations.
This way of constructing the quartets is substantially different from that of Refs.\cite{sasat1,sasaj0} where we have instead adopted a $dynamical$ definition of the $T=0$, $J=0$ quartets. That is to say, in these works, quartets have been constructed independently for each nucleus (via an iterative variational procedure) and no connection whatsoever could be found among quartets of different nuclei. In the present work, owing to the fact that various $T$,$J$ quartets are being considered, this way of proceeding would become prohibitively cumbersome. However, for comparison, we will also carry out an analysis of the ground states of self-conjugate nuclei by applying the same variational procedure of Refs. 
\cite{sasat1,sasaj0}.

The criterion adopted for the selection of the quartets has been that of choosing, as representative of the quartets with a given isospin $T$, those describing the lowest levels with that isospin in nuclei with four active particles outside the inert core of reference. For applications within the $sd$ shell, the inert core is represented by $^{16}$O and the nuclei which have therefore been considered for the definition of the quartets are $^{20}_{10}$Ne$_{10}$, $^{20}_{9}$F$_{11}$ and $^{20}_{8}$O$_{12}$. The lowest states of these nuclei are characterized by $T=$0, 1 and 2, respectively.  Each of these states therefore identifies a quartet with given $T$,$J$ (and a projection $T_z=(N-Z)/2$). We have carried out shell model calculations for these three nuclei and selected the quartets associated with the lowest states. More precisely, we have retained the lowest 6 states in $^{20}$Ne ($0\leq J\leq 6$), 5 states in $^{20}$F ($1\leq J\leq 5$) and 9 states in $^{20}$O ($0\leq J\leq 4$). For all calculations in the $sd$ shell we have employed the USDB interaction \cite{usdb} which gives a pretty good description of the above nuclei. The quartets so selected have then been employed in all the configuration interaction calculations relative to the $sd$ shell.

\begin{figure}[h,b]
\begin{center}
\includegraphics[width=3in,height=3.5in,angle=-90]{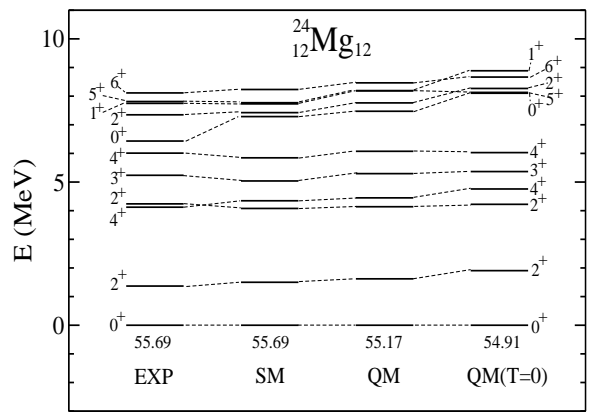}
\caption{
The low-energy spectrum of $^{24}$Mg obtained in the quartet model (QM) compared to the
experimental data (EXP) and to the shell model (SM) results. The spectrum indicated
as QM(T=0) corresponds to a calculation done only with T=0 quartets. In this figure as well as in all the remaining ones relative to nuclei in the $sd$ shell, experimental levels and SM results have been extracted from Ref. \cite{brown}. The number below each spectrum gives the ground state correlation energy, namely the difference between the total ground state energy and the energy in the absence of interaction.}
\label{fig:24mg}
\end{center}
\end{figure}

We start our analysis by examining the self-conjugate nucleus $^{24}$Mg. In Fig. 1, we compare the experimental spectrum of this nucleus with that resulting from a shell model (SM) calculation and with the  spectrum obtained in the quartet model (QM)  when all the selected $T$=0, 1, 2 quartets are taken into account. 
The quartet approach is seen to reproduce well both the SM ground state correlation energy and the SM excited states up to an 
energy of about 9 MeV. The SM is in turn able to fit well the experimental levels. 

Having verified that the selected sets of $T$=0, 1, 2 quartets are sufficient to describe the low-energy spectrum of $^{24}$Mg, it becomes of interest to investigate the role of the different quartets. 
We begin by focusing on isospin. In Fig. 1, on the right hand side, we show the theoretical spectrum obtained in the quartet formalism when only $T$=0 quartets are retained. One can see that this spectrum does not exhibit relevant differences with respect to the full QM calculation. This provides a clear evidence of the marginal role played by the $T$=1 and $T$=2 quartets in the structure of these states. 
As a next step, we concentrate on the ground state by employing only $T=0$  quartets. In Fig. 2, we show how the error in the correlation energy of this state, relative to the SM result, 
varies by reducing, one quartet at a time and starting from the highest one in energy, the set of $T$=0 quartets. The correlation energy remains basically unchanged up to the point where only the lowest $J$=0,2,4 quartets are left. From this point on further reductions in the set of quartets lead to significant variations in the energy. Thus,  these calculations indicate that the $T$=0 quartets with  $J$=0,2,4 play a major role
in the ground state of $^{24}$Mg. Among these quartets, the $T$=0, $J$=0 quartet is by far the one which contributes most to the correlation energy since, as it can be seen in Fig. 2, an approximation in terms of only this quartet accounts for about 94$\%$ of the total energy. 

\begin{figure}[h,t]
\begin{center}
\includegraphics[width=3in,height=3.5in,angle=-90]{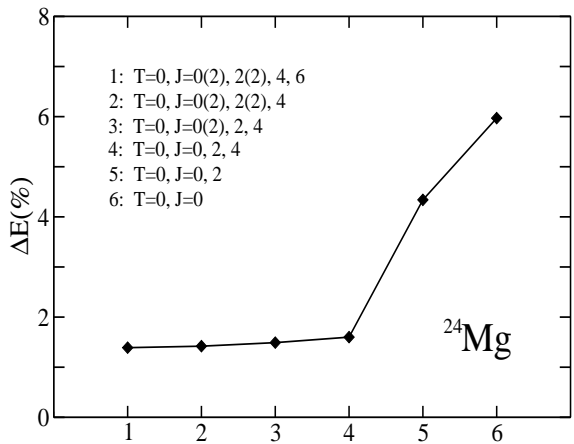}
\caption{
Relative errors (with respect to the SM value) in the ground state correlation 
energy of $^{24}$Mg  calculated in the QM approach
with the quartets indicated in the figure.}
\end{center}
\end{figure}

In this connection, 
a few considerations are needed. The static formulation of the quartets adopted in this work is by construction not the most 
performing one since it does not guarantee a minimum in the ground state energy. The best description of the ground state of 
$^{24}$Mg as a product of two $T=0$, $J=0$ quartets is expected to be obtained by letting these quartets to be distinct from 
one another and by constructing them variationally. An effective way to do that is provided by the procedure adopted in 
Refs. \cite{sasat1,sasaj0}. By applying this procedure one is able to generate a ground state correlation energy which exhausts 95.6$\%$ of 
the exact value. 
This calculation therefore further enhances the role of $T=0$, $J=0$ quartets in this ground 
state and also suggests that the minimum set of $T$=0, $J$=0,2,4 quartets emerged from the above analysis might be
further reduced with a more effective choice of the quartets.

We have searched for a confirmation of the results just discussed by examining another self-conjugate nucleus in the $sd$ shell: $^{28}$Si. We find that for this isotope the set of $T$=0, $J$=0,2,4 quartets already exhaust almost 99$\%$ of the correlation energy. Also in this case, the $T=0$, $J=0$ quartet is found to play a leading role  being able to account by itself for 93.4$\%$ of the total energy. This percentage increases to 98.2$\%$ when the quartets are determined variationally.

So far we have examined only self-conjugate nuclei. The range of nuclei which are accessible in terms of a set of $T$=0,1,2 quartets is, however, much broader. Limiting ourselves to the case of 8 active particles (two quartets) outside the $^{16}$O core, the whole isobaric chain of nuclei with $A$=24 is within reach. We have investigated to what extent the spectra of these nuclei can be described in terms of the full set of $T$=0,1,2 quartets employed in the case of $^{24}$Mg. The results are shown in Fig. 3 where we compare
the low-energy spectra of $^{24}_{11}$Na$_{13}$, $^{24}_{10}$Ne$_{14}$,
$^{24}_{9}$F$_{15}$  and $^{24}_{8}$O$_{16}$ obtained in the quartet model with the SM results and with the experimental data. As one can see, for all these nuclei the quartet formalism
generates spectra which agree well with the SM ones.
It is worth stressing that for the nuclei shown in Fig. 3 the quartets are 
built not only by two protons and two neutrons, as in the case of $N=Z$ nuclei, but also by one proton and three neutrons and by four neutrons. The case of quartets built by four like-particles had been already discussed in Ref. \cite{sasa_jpg} in relation with a treatment of the pairing Hamiltonian.

 \begin{figure*}[h,t]
\centering
\begin{tabular}{cc}
\epsfig{file=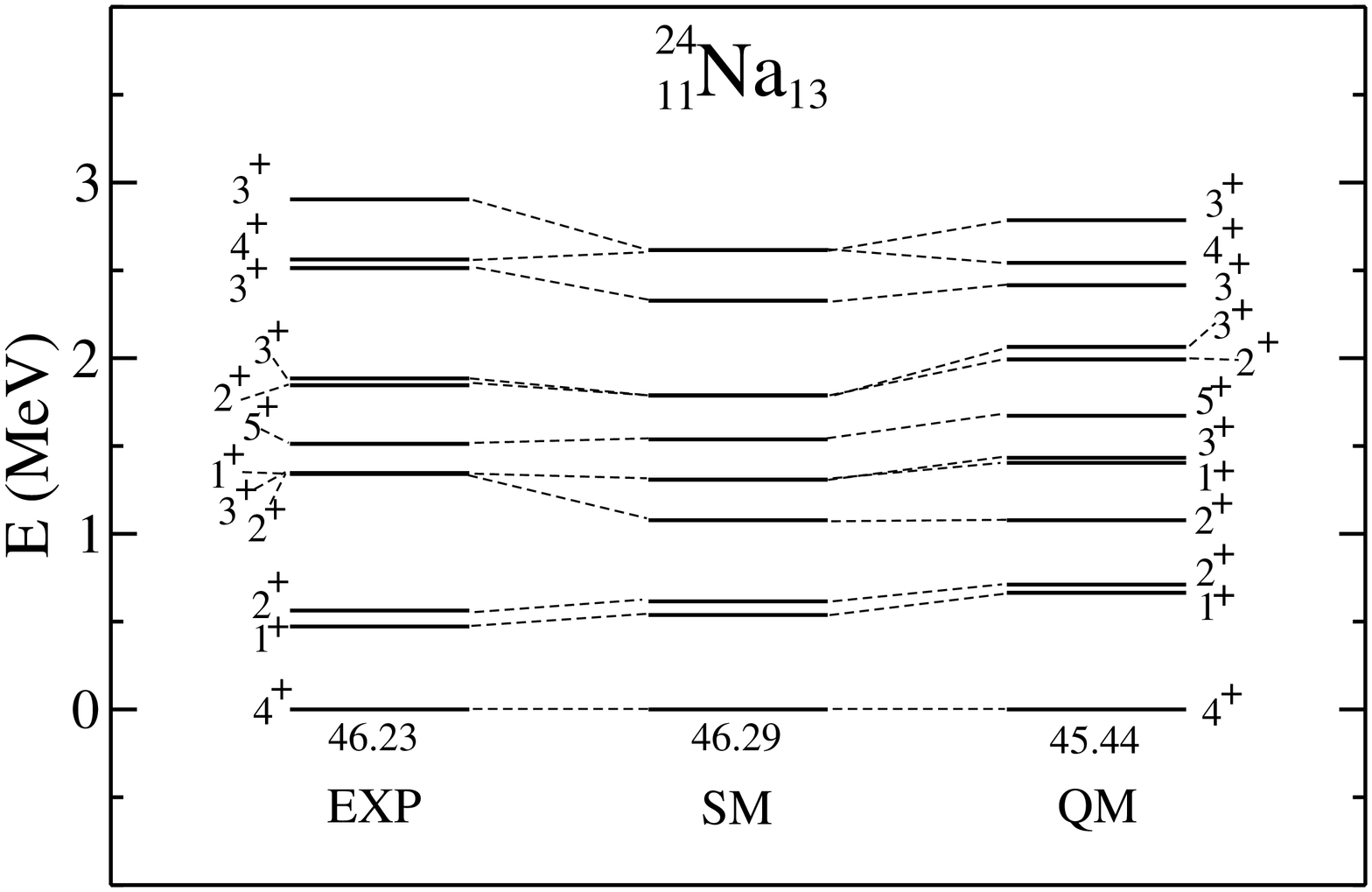,width=3in,height=3.5in,angle=-90} &
\epsfig{file=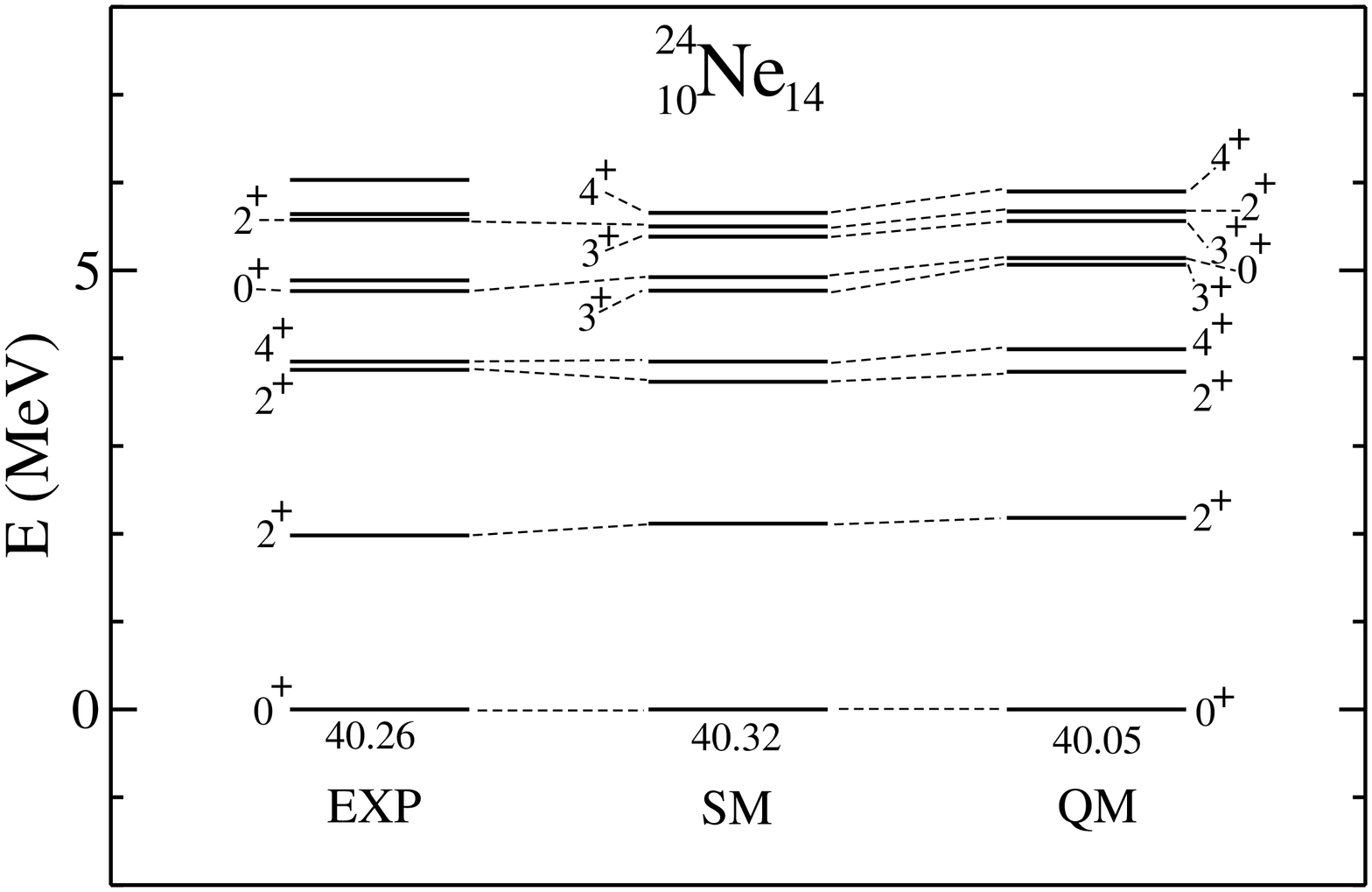,width=3in,height=3.5in,angle=-90} \\
\end{tabular}
\label{fig:contour}
\end{figure*}
 \begin{figure*}[t]
\centering
\begin{tabular}{cc}
\epsfig{file=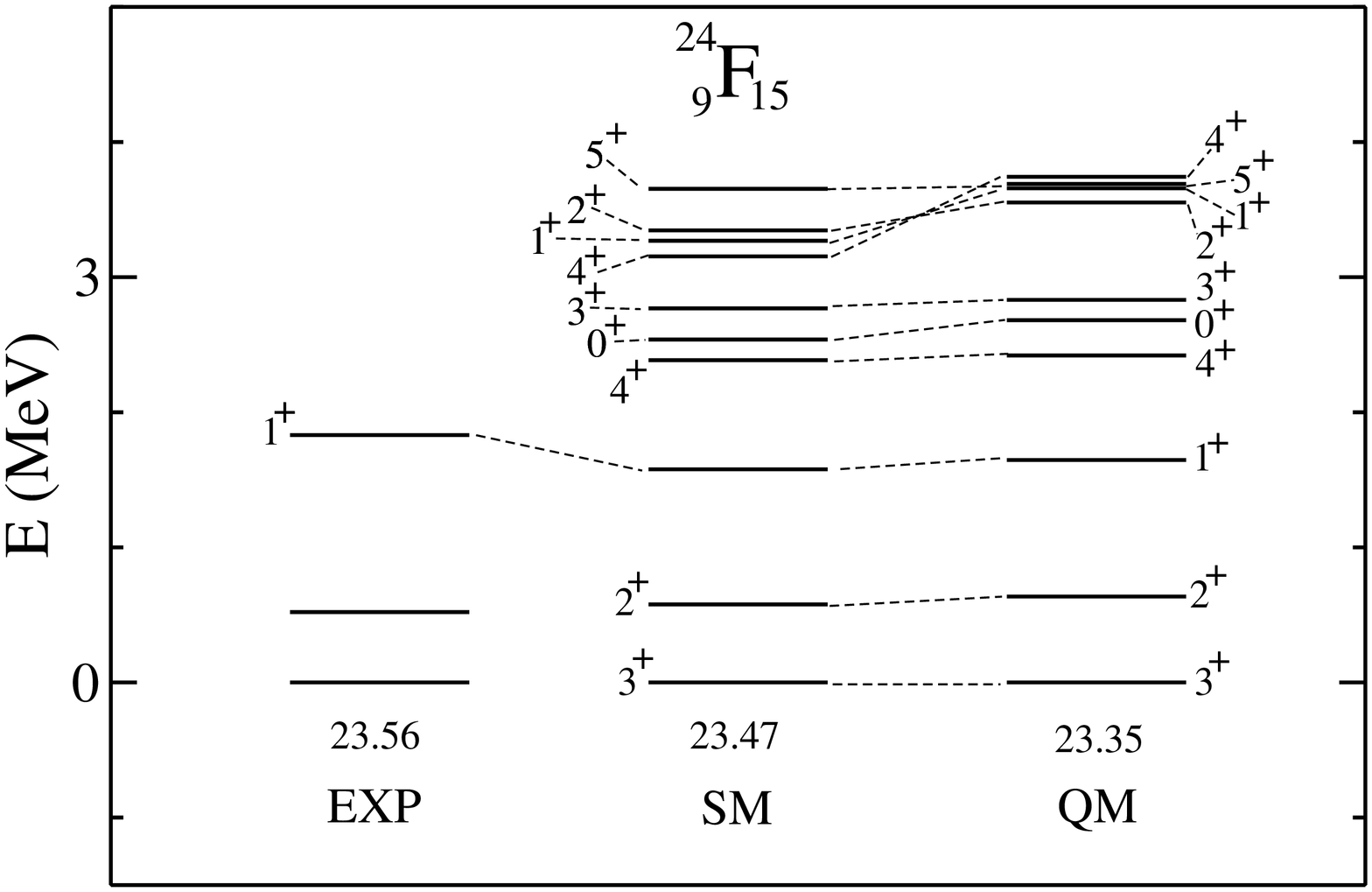,width=3in,height=3.5in,angle=-90} &
\epsfig{file=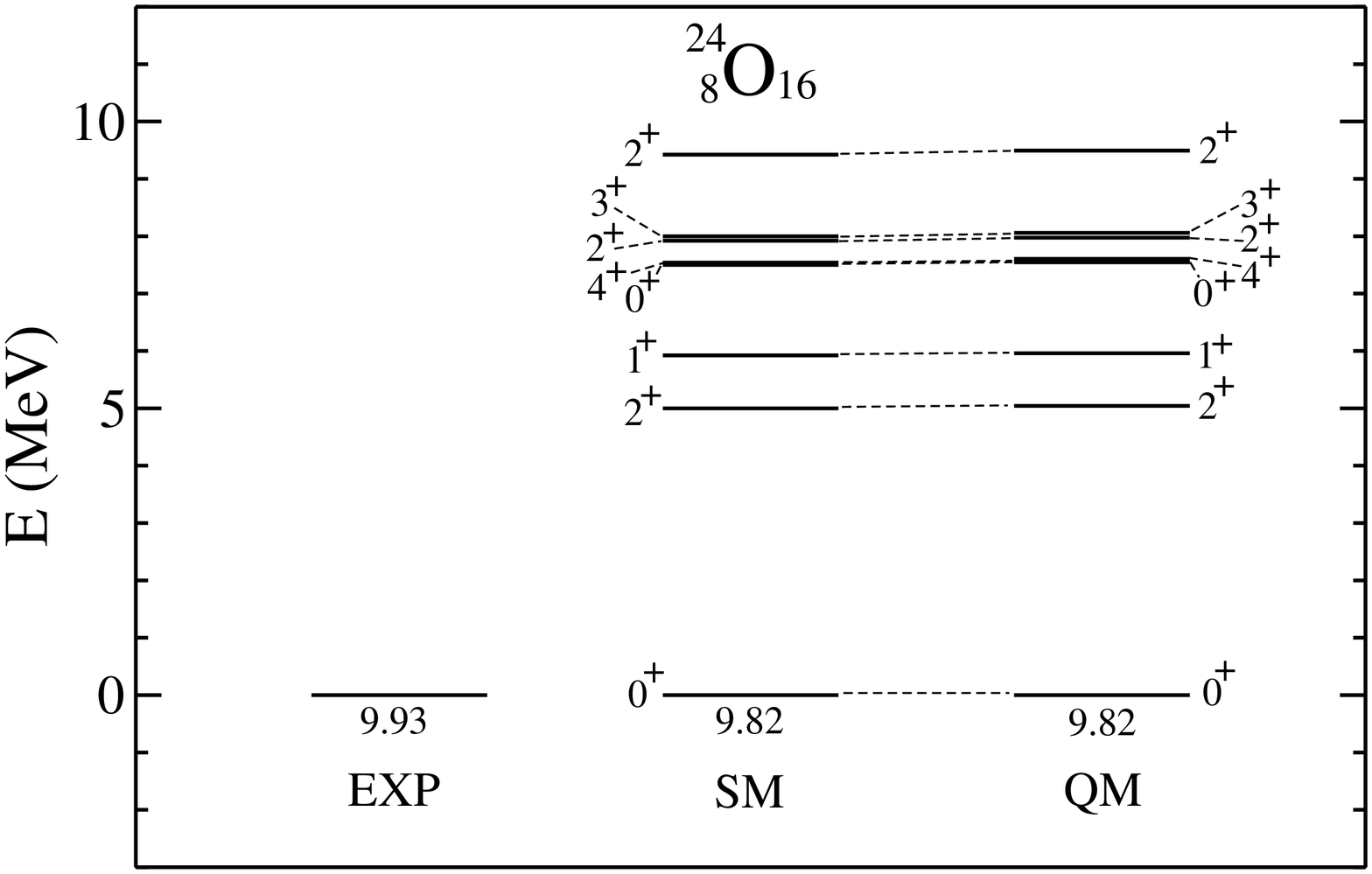,width=3in,height=3.5in,angle=-90} \\
\end{tabular}
\caption{
The spectra of A=24 nuclei generated by the QM approach compared with the
experimental data and the SM predictions. See the caption of Fig. 1 for further details.}
\label{fig:contour}
\end{figure*}

 In order to verify that the results presented so far are not specific to $sd$-shell nuclei,
 we have carried out a similar analysis for the proton-rich nucleus $^{92}_{46}$Pd$_{46}$. Calculations have been done in a space spanned by the $p_{1/2}$, $g_{9/2}$ orbitals using the F-FIT interaction by Johnstone and Skouras \cite{skouras}. In Fig. 4, one sees the low-lying yrast spectrum that is obtained in the QM approach by employing only $T=0$ quartets. These quartets are those associated with the lowest levels of $^{96}_{48}$Cd$_{48}$ (7 levels with $0\leq J\leq 8$) and include also a negative parity level with $J=5$. The QM yrast spectrum is seen to agree well with the experimental one \cite{cederval} as well as with the SM spectrum generated within the same model space by Herndl-Brown \cite{herndl}.
 As in the cases of the $sd$ shell nuclei analyzed above, the $T=0$, $J=0$ quartet is found to play a leading role in the ground state, 
 accounting for almost 99$\%$ of the correlation energy. More details about the structure of $^{92}$Pd in the quartet model 
 will be presented in a separate study.

By summarizing, in  this work we have presented an analysis of $4n$ nuclei in a formalism of quartets, i.e., 
four-body correlated structures characterized by total isospin $T$ and total angular momentum $J$. 
The analysis has concerned the whole isobaric chain of $A=24$ nuclei, therefore ranging from even-even to odd-odd nuclei as well as from self-conjugate nuclei to nuclei with only neutrons in the valence shell.
 For all these nuclei, as well as for the proton-rich nucleus $^{92}$Pd, the quartet formalism has 
 provided a description of the low-energy spectra comparable in accuracy with that of shell model calculations.
 This fact confirms the importance of quartet degrees of freedom in any type  of $4n$ nuclei and validates the present quartet formalism 
 as the appropriate tool for treating them.
 
  \begin{figure}[h,t]
\begin{center}
\includegraphics[width=3in,height=3.5in,angle=-90]{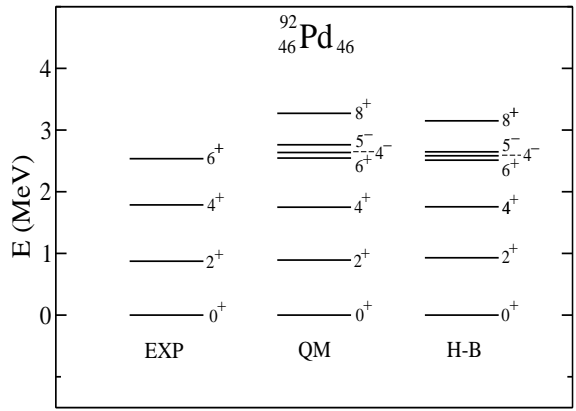}
\caption{
The low-energy yrast spectrum  of $^{92}$Pd obtained in the QM approach compared to the experimental data \cite{cederval} and the
shell model  calculation (H-B) of Ref. \cite{herndl}.}
\label{fig:pd92}
\end{center}
\end{figure}
 
 As a concluding remark, we notice that the description of $4n$ self-conjugate nuclei in terms of T=0 quartets of low angular momenta (J=0,2,4) which has emerged from the present analysis exhibits a striking analogy with that of nuclear collective spectra in a formalism of $S$ ($J$=0), $D$ ($J$=2) and  $G$ ($J$=4) pairs (e.g., see Ref.\cite{cpa}; for a more general 
 shell-model like formalism based on collective pairs see Ref. \cite{msm}). 
 Such an analogy  encourages the extension to quartets of boson mapping
 techniques developed for a microscopic analysis of the Interacting Boson Model (IBM) \cite{ibm}. 
 Thus, if the $s$ ($J=0$) and $d$ ($J=2$) IBM bosons are interpreted as
the images of, respectively, the $T=0$, $J=0$ and $T=0$, $J=2$ quartets,
and to the extent that
the description of the fermionic Hamiltonian in the space of these
quartets can be transferred onto that of a two-body hermitian $sd$ boson
Hamiltonian, with its parameters  eventually absorbing
the contribution of other $T=0$ quartets (if any), then the use of
IBM-type Hamiltonians for the treatment of even-even self-conjugate nuclei
finds a microscopic justification in our analysis.
To our knowledge, in literature there exists only one (old) example of such an use which was carried out, however, on a pure phenomenological basis \cite{duke}.

\vskip 0.3cm

{\it Acknowledgments} 
This work was supported by the Romanian Ministry of Education and Research
through the grant Idei nr 57.


\begin{thebibliography}{10}
\bibitem{ikeda} K. Ikeda, T. Marumori, R. Tamagaki, and H. Tanaka, Prog. Theor. Phys. Suppl. {\bf 52}, 1 (1972).
\bibitem{hoyle}A. Tohsaki, H. Horiuchi, P. Schuck, and G. Ropke,  Phys. Rev. Lett. {\bf 87}, 192501 (2001).
\bibitem{flowers}
B. H. Flowers and M. Vujicic, Nucl. Phys. {\bf 49}, 586 (1963).
\bibitem{arima} A. Arima and V. Gillet, Ann. of Phys. {\bf 66}, 117 (1971). 
\bibitem{yamamura}
J. Eichler and M. Yamamura, Nucl. Phys. A {\bf 182}, 33 (1972).
\bibitem{hasegawa}
M. Hasegawa, S. Tazaki, and R. Okamoto, Nucl. Phys. A{\bf 592}, 45 (1995).
\bibitem{dobes}
 J. Dobes and S. Pittel, Phys. Rev. C {\bf 57}, 688 (1998).
 \bibitem{zelevinsky}
R. A. Senk'ov and V. Zelevinsky, Phys. At. Nucl. {\bf 74}, 1267 (2011).
\bibitem{qcm}
N. Sandulescu, D. Negrea, J. Dukelsky, C. W. Johnson, Phys. Rev. C 
{\bf 85},061303(R) (2012);
N. Sandulescu, D. Negrea, C. W. Johnson, Phys. Rev. C {\bf 86}, 041302 (R) (2012); 
D. Negrea and N. Sandulescu, Phys. Rev. C{\bf 90}, 024322 (2014).
\bibitem{sasat1}
M. Sambataro and N. Sandulescu, Phys. Rev. C {\bf 88}, 061303(R) (2013).
\bibitem{sasaj0}
M. Sambataro, N. Sandulescu, and C.W. Johnson, Phys. Lett. B{\bf740}, 137 (2015).
\bibitem{usdb}B.A. Brown and W.A. Richter, Phys. Rev. C {\bf 74}, 034315 (2006).
\bibitem{brown}www.nscl.msu.edu/~brown/resources/resources.html.
\bibitem{sasa_jpg}M. Sambataro and N. Sandulescu, J. Phys. G: Nucl. Part. Phys. {\bf 40}, 055107 (2013).
\bibitem{skouras} I.P. Johnstone and L.D. Skouras, Eur. Phys. J. A {\bf 11}, 125 (2001).
\bibitem{cederval} B. Cederval et al, Nature {\bf 469}, 68 (2011).
\bibitem{herndl} H. Herndl and B.A. Brown, Nucl. Phys. A {\bf 627}, 35 (1997).
\bibitem{msm}
R. J. Liotta and C. Pomar, Nucl. Phys. A {\bf 362}, 137 (1981).
\bibitem{cpa} F. Catara, A. Insolia, M. Sambataro, E. Maglione, and A. Vitturi, Phys. Rev. C{\bf 32}, 634 (1985).
\bibitem{ibm} F. Iachello and A. Arima, {\it The Interacting Boson Model} (Cambridge University Press, 1987).
\bibitem{duke}
J. Dukelsky, P. Federman, R. P. J. Perazzo, and H. M. Sofia, Phys. Lett. B {\bf 115}, 359 (1982). 

\end{thebibliography}
\end{document}